\documentclass{PoS}

\title{Strongly coupled Graphene on the Lattice}

\ShortTitle{Strongly coupled Graphene on the Lattice}

\author{\speaker{Timo A. L\"ahde} \\
       Helsinki Institute of Physics and Department of Applied Physics, 
       Aalto University, FI-00076 Aalto, Espoo, Finland \\
       E-mail: \email{talahde@gmail.com}}

\author{Joaqu\'\i n E. Drut \\
       Theoretical Division, Los Alamos National Laboratory, Los Alamos, NM 87545-0001, USA \\
       E-mail: \email{joaquindrut@gmail.com}}

\abstract{The two-dimensional carbon allotrope graphene has recently attracted a lot of attention from researchers
in the disciplines of Lattice Field Theory, Lattice QCD and Monte Carlo calculations. This interest has been prompted
by several remarkable properties of the conduction electrons in graphene. For instance, 
the conical band structure of graphene at low energies is strongly reminiscent of relativistic Dirac fermions.
Also, due the low Fermi velocity of $v_F^{} \simeq c/300$, where $c$ is the speed of light in vacuum, the
physics of the conduction electrons in graphene is qualitatively similar to Quantum Electrodynamics in a 
strongly coupled regime. In turn, this opens up the prospect of the experimental realization of gapped, 
strongly correlated states in the electronic phase diagram of graphene. Here, we review the experimental
and theoretical motivations for Lattice Field Theory studies of graphene, and describe the directions that
such research is likely to progress in during the next few years. We also give a brief overview of the two main
lattice theories of graphene, the hexagonal Hubbard theory and the low-energy Dirac theory. Finally, we
describe the prospect of extracting response functions, such as the electric conductivity, using Lattice
Field Theory calculations.
}

\FullConference{XXIX International Symposium on Lattice Field Theory \\
		 July 10 -- 16 2011 \\
		 Squaw Valley, Lake Tahoe, California}

\begin{document}


\section{Introduction}

Graphene is a novel two-dimensional carbon nanomaterial with unusual electronic properties. These deviate from
the standard theory of conduction due to the linear dispersion of the charge carriers~\cite{graphene_review}. 
In spite of the ``relativistic'' 
energy-momentum relation, the electrons in graphene move non-relativistically. This is evidenced by the low
Fermi velocity of $v_F^{} \simeq c/300$, where $c$ denotes the speed of light in vacuum.
While the band structure of graphene is classified as semimetallic, the material conducts 
electricity better than silver and outperforms the carrier mobility in silicon due to its high resistance to impurities and chemical 
damage. Most importantly, graphene is inspiring technological innovations and improvements,
such as graphene-based transistors and integrated 
circuits, transparent conducting electrodes, solar cells and ultracapacitors. The discovery of this revolutionary material 
in 2004 was awarded the Nobel Prize in physics in 2011.

Compared to Quantum Electrodynamics (QED), the
electron-electron coupling $\alpha_g^{}$ in graphene is enhanced by a factor of $c/v_F^{}$, and is thus roughly 300~times larger
than the fine-structure constant $\alpha \simeq 1/137$~\cite{Herbut}. 
Moreover, as the screening length diverges at the neutral point, 
the Landau picture which admits the mapping of a strongly interacting electronic liquid onto a gas of non-interacting fermions
breaks down. In this situation, the electrical conductivity is of particular interest, as it is
under intense study in suspended graphene, which can be
routinely manufactured and electrically contacted. The temperature- and frequency dependence of the 
conductivity is also a central design parameter in nanoelectronic applications, such as transistors and current switches.


Graphene, as a strongly interacting quantum mechanical many-body system, ranks among the most challenging problems in 
theoretical and computational physics to date. Other prominent examples where complex behavior emerges in many-body
systems at strong coupling
include, to name a few, high-$T_c^{}$ superconductivity, superfluidity in ultracold atomic gases, and the formation of hadrons in QCD.
While significant progress 
has been made in the theoretical sector using analytical approaches (such as perturbative, renormalization group
and self-consistent treatments), non-perturbative methods which take full account of quantum 
mechanical fluctuations offer the best prospect for a complete, {\it a priori} solution. A great driving force behind
our understanding of strongly coupled quantum systems is Lattice QCD, whereby the properties of QCD are computed from 
first principles on a discretized space-time lattice. Due to advances in algorithmics as well as in computer power, Lattice QCD is 
able to predict the properties of hadrons (such as the masses of baryonic resonances) to percent-level accuracy. 
The discovery of graphene has also promoted the adoption of Lattice QCD methods, in particular the Hybrid Monte Carlo (HMC)
algorithm, in condensed matter physics~\cite{Graphene_DL,Graphene_Hands,Brower_graphene,Araki}. In turn, the HMC algorithm has 
lead to great
progress in the closely related area of Hubbard-like models in atomic physics, in particular for the strongly interacting
Fermi gases in the unitarity limit~\cite{Contact_DL}.




\section{Graphene at strong electron-electron coupling}

Graphene is often described both as a strongly interacting many-fermion system, and as an 
``ultimate system of non-interacting electrons''. Recent experiments with suspended graphene
have provided increasing evidence for the ``missing'' electron-electron interaction. While the
spontaneous breakdown of the semimetallic phase is yet to be detected, observed phenomena include the fractional quantum Hall 
effect~\cite{Bolotin_FQHE} in an external magnetic field, and a strong upward renormalization of the Fermi velocity
at the neutral point~\cite{Elias_etal}. Also, a resistive state in a suspended graphene bilayer has recently been 
reported at zero magnetic field~\cite{bilayer_gap}, and hypothesized to be of interaction-driven origin.
Taken together, these developments suggest that an investigation of emergent strong-coupling phenomena in graphene is timely,
which should also clarify the question whether the electron-electron interaction in graphene is key to our understanding of the 
electronic phase diagram and transport, or whether such effects can safely be neglected. The combination of Lattice Field Theory
and MC calculations, formulated either in terms of a hexagonal Hubbard model based on the tight-binding description of
graphene, or in terms of a low-energy theory of Dirac fermions, is a promising way to proceed as it captures the physics of 
strong electron-electron interaction without uncontrolled approximations. Below is a (by no means exclusive) list of objectives
that are likely to be central in the application of Lattice Field Theory to graphene in the near future:

\begin{itemize}

	\item {\bf The zero-temperature electronic phase diagram of graphene.} 
	What is the critical value of the electron-electron coupling $\alpha_g^{}$ where the 
	semimetallic properties of graphene break down? Is suspended graphene above or below this critical coupling at zero
	temperature? Is it possible to induce an insulating state by a technologically feasible amount
	of strain (increasing the lattice constant of graphene by a few tens of percent) to reduce the probability for the electrons 
	to tunnel between neighboring carbon atoms? What are the
	critical exponents of the semimetal-insulator transition? How is the Fermi velocity $v_F^{}$ renormalized by interactions? 
	To what extent is the critical coupling and the associated exponents affected by 
	screening of the electron-electron interaction? 
	
	\item {\bf Finite temperature and gapped phases in graphene.} 
	The effects of finite temperature on gapped strong-coupling phases in graphene
	are especially interesting, as the Mermin-Wagner theorem prohibits spontaneous breaking of a continuous symmetry
	in a two-dimensional system. However, there is considerable evidence from MC studies
	by Hands {\it et al.}~\cite{Graphene_Hands} that 
	suspended graphene undergoes
	a Kosterlitz-Thouless (KT) transition into a gapped phase at a finite temperature, which is currently poorly known. How is the 
	zero-temperature phase diagram of graphene
	reflected in the finite-temperature properties? What is the critical temperature of the putative KT transition?

	\item {\bf The electrical conductivity of graphene.} How does the electron-electron interaction affect the conductivity 
	$\sigma(\omega,T)$ as a function of frequency $\omega$ and temperature $T$? To what extent 
	is the minimal DC conductivity $\sigma(\omega=0,T)$ determined by the electron-electron interaction, and how
	can the measured (see Ref.~\cite{Elias_etal}) temperature dependence of the minimal conductivity at low~$T$ 
	be explained? How close is suspended graphene (and graphene on a dielectric substrate) to a 
	quantum critical point where the semimetallic phase breaks down? 
					 	
	\item {\bf The viscosity of the electrons in graphene.} 
	Theoretical estimates~\cite{Fritz_viscosity} of the ratio of shear viscosity to entropy 
	density $\eta/s$ have put forward the possibility that the viscosity in graphene is unusually small, which was interpreted 
	as an indicator of complex fluid dynamics, observable on length scales of $\simeq 1~\mu$m.
	What is the magnitude and temperature dependence of the viscosity in the presence of interactions? 
	How does the appearance of spontaneously gapped phases affect the viscosity? 
	Within a MC calculation, can sufficient statistics be obtained to find a clear signal for the viscosity?
	
	\item {\bf Superfluidity in a graphene bilayer.} 
	In a bilayer with equal populations of electrons and holes on the
	respective layers, the attractive electron-hole interaction induces pair formation of electrons and holes. 
	What is the critical temperature $T_c^{}$ for superfluidity of electron-hole pairs in a such a graphene bilayer? 
	Depending on the amount of screening, $T_c^{}$ has been suggested 
	to be as high as $300$~K, while $T_c^{} \simeq \mu$K in a 
	more pessimistic scenario~\cite{Bilayer}. If the system is imbalanced (more electrons than holes or {\it vice versa}), 
	$T_c^{}$ will decrease. What is the critical imbalance where $T_c^{}$ reaches zero?
			
	\item {\bf Ultracold fermionic atoms in a hexagonal optical lattice.} What similarities exist between the electronic 
	properties of graphene and the analogous observables of ultracold fermionic atoms (such as potassium-40) confined 
	to a hexagonal optical lattice? 
	The tunneling amplitude and the interparticle coupling can be controlled over a wide range on an optical lattice, 
	the latter using the Feshbach resonance technique~\cite{optical_lattice}. In real graphene, such tuning can in principle be
	achieved using tension and dielectric materials, albeit in a much more cumbersome and limited way. 
	Significantly, the interparticle coupling on optical lattices can be increased beyond $\alpha_g^{} \simeq 2$, providing access to 
	physics that may only occur in an ``unphysical'' parameter range in real graphene. 
	
\end{itemize}


\section{Hubbard theory of graphene}

Graphene can be described as a tight-binding model with (non-local) electron-electron couplings.
The corresponding (non-relativistic) Hamiltonian is
\begin{eqnarray}
\hat H \!\!\!&=&\!\!\! 
-t \sum_{\langle \textbf{\small i},\textbf{\small j} \rangle,\sigma} 
\psi_{\textbf{\small i},\sigma}^\dagger \psi_{\textbf{\small j},\sigma}^{}
-t' \!\! \sum_{\langle\langle \textbf{\small i},\textbf{\small j} \rangle\rangle,\sigma} 
\psi_{\textbf{\small i},\sigma}^\dagger \psi_{\textbf{\small j},\sigma}^{}
+ U_{00}^{} \sum_{\textbf{\small i}} n_{\textbf{\small i},\uparrow}^{} n_{\textbf{\small i},\downarrow}^{}
+ \frac{1}{2} \sum_{\textbf{\small i} \neq \textbf{\small j}, \sigma,\sigma'} U_{\textbf{\small ij}}^{} \,
n_{\textbf{\small i},\sigma}^{} n_{\textbf{\small j},\sigma'}^{},
\label{QMC}
\end{eqnarray}
where $\psi_{\textbf{\small i},\sigma}$ annihilates an electron at lattice site $\textbf{\small i}$ with 
spin $\sigma \!\in\! \{\uparrow,\downarrow\}$ and $n_{\textbf{\small i},\alpha}^{} \!\!\equiv
\psi_{\textbf{\small i},\alpha}^\dagger \psi_{\textbf{\small i},\alpha}^{}$.
The tunneling amplitudes between nearest- and next-to-nearest neighbor sites are $t \simeq 2.5$~eV 
and $t' \simeq 0.1$~eV, respectively. If only the on-site repulsion $U_{00}^{}$ is accounted for,
the phase diagram of Eq.~(\ref{QMC}) has been computed using the Quantum Monte Carlo (QMC) approach of
Ref.~\cite{QMC_Meng}.
In such a treatment, graphene exhibits a semimetallic phase~(SM) at weak coupling, a gapped spin-liquid phase~(SL) 
at intermediate couplings and an antiferromagnetic Mott insulator~(AFMI) at strong coupling. As $U_{00}^{}/t \simeq 3.3$ in 
suspended graphene, such systems are tantalizingly close to the critical coupling $U_{00}^{}/t \simeq 3.5$ for a 
semimetal-insulator transition into the SL phase.

At least two mechanisms exist which can change this picture. First, $U_{00}^{}/t$ increases under the application of strain, due to
the decreased overlap of the carbon orbitals. A technologically feasible amount of strain, increasing the lattice constant of 
graphene by a few percent, may then suffice to induce the SL state. Second, the non-local couplings in Eq.~(\ref{QMC}) should
not be neglected at low densities, where the electron-electron interaction remains unscreened. Indeed, the couplings
$U_{\textbf{\small ij}}$ have been computed in density functional theory~(DFT) by Ref.~\cite{Wehling}, 
which yielded $U_{01}^{}/t \simeq 2.0$ with further sizeable contributions at longer ranges. A particularly interesting question
is in what way such non-local couplings affect the electronic phase diagram, in particular the critical coupling for the SL state.
Notably, the Hybrid Monte Carlo~(HMC) algorithm~\cite{HMC}
which enables global lattice updates, has recently been successfully
employed for systems with a Hamiltonian similar to Eq.~(\ref{QMC}), in particular the Unitary Fermi Gas~(UFG).


\section{Dirac theory of graphene}

Relativistic QFT can be applied to graphene, as the electronic dispersion relation
becomes linear in the vicinity of the neutral point~\cite{Herbut}.
The resulting linearized low-energy theory, valid in the vicinity of the ``Dirac points'' of graphene,
can be studied at strong electron-electron interaction using the Lattice Monte Carlo~(LMC) framework, and complements the 
abovementioned Hubbard approach. As in Lattice QCD and Lattice Field Theory in general, Euclidean space-time is used in 
order to obtain a positive definite probability measure.
The resulting low-energy QFT of graphene can then be formulated in terms of the action
\begin{equation}
S_E^{} \equiv \frac{1}{2g^2_{}} \int d^3x \, dt \, (\partial_i^{} A_0^{})^2_{}
- \sum_{a=1}^{N_f^{}} \int d^2x \, dt \, \bar\psi_a^{} D[A_0^{}] \psi_a^{}, 
\quad D[A_0^{}] = \gamma_0^{}(\partial_0^{} + iA_0^{}) + v_F^{}\gamma_i^{}\partial_i^{}, 
\label{EFT}
\end{equation}
where $\psi$ (with $\bar\psi \equiv \psi^\dagger \gamma_0^{}$)
is a four-component fermion field, and $\gamma_\mu^{}$ denotes the Dirac gamma matrices.
A graphene monolayer corresponds to $N_f^{} \!=\! 2$, while a bilayer is described by $N_f^{} \!=\! 4$ in the absence
of interlayer couplings. Interlayer tunneling can be accounted for by introducing two coupled
monolayers, giving rise to the characteristic quadratic dispersion of graphene bilayers. 
The electron-electron interaction is mediated by the instantaneous gauge field $A_0^{}$, with
coupling constant $g^2 \equiv e^2/\epsilon$. Here $\epsilon$ is the effective permittivity of the medium (for example a hemisphere of 
SiO$_2^{}$ substrate). The strength of the electron-electron
interaction is controlled by the ``fine structure constant'' $\alpha_g^{} \equiv e^2/(4\pi\epsilon v_F^{})$ (in units where $\hbar = c = 1$),
such that low values of $v_F^{} \simeq 1/300$ and $\epsilon \simeq 1$ translate into a large electron-electron coupling. In the 
presence of a dielectric substrate, $\epsilon > 1$ due to the increased permittivity. It should be noted that Eq.~(\ref{EFT}) is an
example of a ``reduced gauge theory'', as the gauge field $A_0^{}$
propagates in one more spatial dimension than the fermion field $\psi$, which is confined to a $(2+1)$-dimensional surface.

The LMC treatment of graphene is based on a discretized version of Eq.~(\ref{EFT}) on a lattice of space-time points.
Unlike the QMC treatment of graphene, this lattice has no correspondence to the ``physical'' hexagonal lattice
of carbon atoms. Instead, physical predictions are obtained in the continuum limit of vanishing lattice spacing. As the continuum
limit can only be approached, a question arises to what extent the space-time lattice
distorts the continuum physics. For this purpose, different discretizations of Eq.~(\ref{EFT}) have been developed, which
share the same continuum limit but emphasize different aspects of the physics on a lattice. In Lattice QCD, a standard choice
is staggered (or Kogut-Susskind) fermions~\cite{staggered}, which only partially respect the symmetries of the continuum theory. 
The use of overlap fermions (see Ref.~\cite{overlap} and references therein), while significantly more computationally expensive, allows 
for a much closer approach to the continuum limit in studies of emergent strong-coupling phenomena, such as spontaneous chiral
symmetry breaking.


\section{Computation of transport properties}

A brief outline of the extraction of the transport properties of graphene using LMC calculations is given here, in order to
provide an idea of the computational intricacies and challenges. An experimentally readily accessible and relevant 
observable is the minimal DC ($\omega = 0$) conductivity $\sigma(n \!=\! 0, T)$, which involves the computation of the 
Euclidean correlator 
\begin{equation}
G(\tau) = \int d^2x \, \langle J^\dagger(\tau,{\textbf x}) J(0,{\textbf 0})\rangle, \quad\quad
J(\tau,{\textbf x}) \,\equiv \sum_{k=1,2}\bar\psi(\tau,{\textbf x}) \gamma_k^{} \psi(\tau,{\textbf x}),
\label{corr}
\end{equation}
where $\rho(\omega)$ is obtained from the correlator $G(\tau)$ (which is measured from the MC data) as
\begin{equation}
G(\tau) = \int_0^\infty \frac{d\omega}{2\pi} \, K(\omega,\tau) \rho(\omega), \quad\quad
K(\omega,\tau) = \frac{\cosh(\omega\tau-\omega/2T)}{\sinh(\omega/2T)}.
\label{inv}
\end{equation}
The minimal conductivity
is then given by the zero-frequency limit of the spectral function $\rho(\omega)$ according to
\begin{equation}
\frac{\sigma(n \!=\! 0,T)}{T} \:=\: \lim_{\omega \to 0} \,\frac{\rho(\omega)}{6\,\omega T},
\end{equation}
where the temperature $T$ is related to the temporal extent $N_\tau$ of the Euclidean space-time lattice by $1/T = aN_\tau$, and
$a$ is the temporal lattice spacing. The inversion of Eq.~(\ref{inv}) is complicated by the fact that $G(\tau)$ is obtained by 
MC calculation at a discrete set of points $\tau_i^{}$, where the number of data points $i$ is typically ${\mathcal O}(10)$ due
to CPU power limitations and algorithm scaling, 
while $\rho(\omega)$ is in principle a continuous function of $\omega$. As the integration range in $\omega$ is discretized into 
$N_\omega^{} \sim {\mathcal O}(10^3)$ points, a straightforward inversion is ill-defined.

The numerical inversion of expressions similar to Eq.~(\ref{inv}) has become
the focus of intense theoretical and computational efforts. Here, we shall focus on the studies at Swansea University of the behavior of
the conductivity in the vicinity of the deconfinement transition in QCD at high temperatures by Aarts {\it et al.}
in Ref.~\cite{Hands_QCD}. These efforts have led to the development of 
a promising algorithm, which allows for a numerically stable calculation of $\rho(\omega)$ 
using a combination of Bayesian analysis and the maximum entropy method (MEM). In the Bayesian approach, one constructs
the ``most probable'' spectral function by minimizing a conditional probability 
$P\left[\rho| DH\right]$, where $D$ denotes the available data on $G(\tau)$ 
and $H$ some additional ``prior knowledge''. In the MEM approach, this additional knowledge is introduced by the
``entropy term''
\begin{equation}
P\left[\rho| DH\right] = \exp\left(-\frac{1}{2}\chi^2 + \alpha S\right), \quad
S = \int_0^\infty \frac{d\omega}{2\pi} \left[
\rho(\omega) - m(\omega) - \rho(\omega) \ln \! \frac{\rho(\omega)}{m(\omega)} \right],
\end{equation}
where $\chi^2$ is the standard likelihood function, and $\alpha$ is a parameter which controls the relative
weight of the data and the prior knowledge, which is introduced through the ``default model'' function $m(\omega)$.
Unfortunately, the MEM analysis of current-current correlators in Lattice QCD suffers from numerical 
instabilities and poor convergence, which make a model-independent determination of $\sigma$ in the
limit $\omega \to 0$ difficult. In this situation, performing the analysis in terms of the new quantities
\begin{equation}
\bar K(\omega,\tau) = \frac{\omega}{2T} \, K(\omega,\tau), \quad\quad
\bar \rho(\omega,\tau) = \frac{2T}{\omega} \, \rho(\omega,\tau), \quad\quad
a^2 \bar m(\omega) = \bar m_0^{} (b + a\omega). 
\label{newstuff}
\end{equation}
was found in Ref.~\cite{Hands_QCD} to circumvent the numerical instabilities at small $\omega$ to a
large extent. 
In the context of Lattice QCD, the zero-frequency conductivity has been shown to be independent 
of $b$ to an accuracy of $10-20\%$. While such accuracy in itself is sufficient to produce novel and valuable information, 
it should also be noted that as far as CPU power is concerned, the situation for graphene is more favorable
than the situation for QCD, due to the lower dimensionality of the graphene problem. This advantage can be translated
to greater MC statistics, but more significantly to a larger extent $N_\tau^{}$ in the temporal lattice dimension, which
will improve the accuracy and model-independence of the MEM analysis.



\begin{thebibliography}{99}

\bibitem{graphene_review}
A.~H.~Castro Neto, F.~Guinea, N.~M.~R.~Peres, K.~S.~Novoselov, and A.~K.~Geim,
Rev.\ Mod.\ Phys. {\bf 81}, 109 (2009).

\bibitem{Herbut}
I.~F.~Herbut,
Phys.\ Rev.\ Lett. {\bf 97}, 146401 (2006);
I.~F.~Herbut, V.~Juri\v{c}i\'c, and O.~Vafek,
Phys.\ Rev.\ B {\bf 80}, 075432 (2009).

\bibitem{Graphene_DL}
J.~E.~Drut and T.~A.~L\"ahde,
Phys.\ Rev.\ Lett. {\bf 102}, 026802 (2009);
{\it ibid.} B {\bf 79}, 165425 (2009);
{\it ibid.} B {\bf 79}, 241405 (2009).

\bibitem{Graphene_Hands}
W.~Armour, S.~Hands, and C.~Strouthos, 
Phys.\ Rev.\ B {\bf 81}, 125105 (2010);
{\it ibid.} B {\bf 84}, 075123 (2011).

\bibitem{Brower_graphene}
R.~C.~Brower, C.~Rebbi, and D.~Schaich,
arXiv:1101.5131.

\bibitem{Araki}
Y.~Araki and T.~Hatsuda,
Phys.\ Rev.\ B {\bf 82}, 121403(R) (2008).

\bibitem{Contact_DL}
J.~E.~Drut, T.~A.~L\"ahde, and T.~Ten,
Phys.\ Rev.\ Lett. {\bf 106}, 205302 (2011).

\bibitem{Bolotin_FQHE} 
X.~Du {\it et al.},
Nature~(London) {\bf 462}, 192 (2009);
K.~I.~Bolotin {\it et al.},
Nature (London) {\bf 462}, 196 (2009).

\bibitem{Elias_etal} 
D.~C.~Elias {\it et al.}, 
Nature\ Phys.\ {\bf 7}, 701 (2011).

\bibitem{bilayer_gap}
J.~Martin {\it et al.},
Phys.\ Rev.\ Lett. {\bf 105}, 256806 (2010).

\bibitem{Fritz_viscosity}
M.~M\"uller, J.~Schmalian, and L.~Fritz,
Phys.\ Rev.\ Lett. {\bf 103}, 025301 (2009).








\bibitem{Bilayer}
H.~Min, R.~Bistritzer, J.-J.~Su, and A.~H.~MacDonald,
Phys.\ Rev.\ B {\bf 78}, 121401 (2008);
M.~Yu.~Kharitonov and K.~Efetov,
Semiconductor Science and Technology {\bf 25}, 034004 (2010).


\bibitem{optical_lattice}
I.~Bloch, J.~Dalibard, and W.~Zwerger,
Rev.\ Mod.\ Phys. {\bf 80}, 885 (2008);
R.~J\"ordens, N.~Strohmaier, K.~G\"unter, H.~Moritz, and T.~Esslinger,
Nature (London) {\bf 455}, 204 (2008).




\bibitem{QMC_Meng}
T.~Paiva, R.~T.~Scalettar, W.~Zheng, R.~R.~P.~Singh, and J.~Oitmaa,
Phys.\ Rev.\ B {\bf 72}, 085123 (2005);
Z.~Y.~Meng, T.~C.~Lang, S.~Wessel, F.~F.~Assaad, and A.~Muramatsu,
Nature (London) {\bf 464}, 847 (2010).

\bibitem{Wehling}
T.~O.~Wehling {\it et al.},
Phys.\ Rev.\ Lett. {\bf 106}, 236805 (2011).




\bibitem{HMC}
S.~Duane, A.~D.~Kennedy, B.~J.~Pendleton, and D.~Roweth,
Phys.\ Lett.\ B {\bf 195}, 216 (1987).

\bibitem{staggered}
C.~Burden and A.~N.~Burkitt,
Europhys.\ Lett.\ {\bf 3}, 545 (1987).

\bibitem{overlap}
Z.~Fodor, S.~D.~Katz, and K.~K.~Szab\'o,
JHEP {\bf 08}, 003 (2004).

\bibitem{Hands_QCD}
G.~Aarts, C.~Allton, J.~Foley, S.~Hands, and S.~Kim,
Phys.\ Rev.\ Lett. {\bf 99}, 022002 (2007); {\it ibid.} Proc.\ Sci.\ {\bf LAT2006}, 134 (2006).




\end{thebibliography}
\end{document}